\documentclass[10pt,pra,twocolumn,showpacs,superscriptaddress,floatfix]{revtex4-2}
\usepackage{graphicx}
\usepackage{amssymb}
\usepackage{fontspec}
\usepackage{amsmath,esint}
\usepackage{amsthm}
\usepackage[colorlinks,
linkcolor=blue,      
anchorcolor=blue, 
citecolor=purple]{hyperref}
\usepackage{color,xcolor}
\usepackage{subfigure}
\usepackage{algpseudocode}
\usepackage{soul}
\usepackage{algorithm}
\usepackage{algorithmicx}
\usepackage{lineno}
\usepackage{soul,xcolor}
\usepackage{color}
\usepackage{graphicx}
\usepackage{dcolumn}
\usepackage{bm}
\usepackage{graphicx}
\usepackage{textcomp,mathcomp}
\usepackage{enumerate}
\usepackage{lineno}
\usepackage{soul}
\usepackage[thicklines]{cancel}
\usepackage{tikz}
\usetikzlibrary{matrix,decorations.pathreplacing,calc}
\newcommand{\ket}[1]{|#1\rangle}
\newcommand{\bra}[1]{\langle#1|}
\newcommand{\Fref}[1]{Fig.\ref{#1}}
\newcommand{\Eref}[1]{Equ.\ref{#1}}
\newcommand{\Tref}[1]{Tab.\ref{#1}}
\usepackage[english]{babel}
\PassOptionsToPackage{english}{babel}
\begin{document}

\title{Discovering autonomous quantum error correction via deep reinforcement learning}

\author{Yue Yin}
\affiliation{Zhiyuan College, Shanghai Jiao Tong University, Shanghai 200240, P.R.  China}

\author{Tailong Xiao}
\email{tailong\_shaw@sjtu.edu.cn}
\affiliation{State Key Laboratory of Photonics and Communications, Institute for Quantum Sensing and Information Processing, Shanghai Jiao Tong University, Shanghai 200240, P.R.  China}
 \affiliation{Hefei National Laboratory, Hefei, 230088, P.R. China}
\affiliation{Shanghai Research Center for Quantum Sciences, Shanghai, 201315, P.R. China}

\author{Xiaoyang Deng}
\affiliation{Zhiyuan College, Shanghai Jiao Tong University, Shanghai 200240, P.R.  China}
\affiliation{State Key Laboratory of Photonics and Communications, Institute for Quantum Sensing and Information Processing, Shanghai Jiao Tong University, Shanghai 200240, P.R.  China}

\author{Ming He}
\affiliation{AI Lab, Lenovo Research, Beijing 100094, P.R.  China}

\author{Jianping Fan}
\affiliation{AI Lab, Lenovo Research, Beijing 100094, P.R.  China}

 \author{Guihua Zeng}
\affiliation{State Key Laboratory of Photonics and Communications, Institute for Quantum Sensing and Information Processing, Shanghai Jiao Tong University, Shanghai 200240, P.R.  China}
 \affiliation{Hefei National Laboratory, Hefei, 230088, P.R. China}
\affiliation{Shanghai Research Center for Quantum Sciences, Shanghai, 201315, P.R. China}

\date{\today}
\begin{abstract}
Quantum error correction is essential for fault-tolerant quantum computing. However, standard methods relying on active measurements may introduce additional errors. Autonomous quantum error correction (AQEC) circumvents this by utilizing engineered dissipation and drives in bosonic systems, but identifying practical encoding remains challenging due to stringent Knill-Laflamme conditions. In this work, we utilize curriculum learning enabled deep reinforcement learning to discover Bosonic codes under approximate AQEC framework to resist both single-photon and double-photon losses.
We present an analytical solution of solving the master equation under approximation conditions, which can significantly accelerate the training process of reinforcement learning. The agent first identifies an encoded subspace surpassing the breakeven point through rapid exploration within a constrained evolutionary time-frame, then strategically fine-tunes its policy to sustain this performance advantage over extended temporal horizons. We find that the two-phase trained agent can discover the optimal set of codewords, i.e., the Fock states $\ket{4}$ and $\ket{7}$ considering the effect of both single-photon and double-photon loss. We identify that the discovered code surpasses the breakeven threshold over a longer evolution time and achieve the state-of-art performance. We also analyze the robustness of the code against the phase damping and amplitude damping noise.
Our work highlights the potential of curriculum learning enabled deep reinforcement learning in discovering the optimal quantum error correct code especially in early fault-tolerant quantum systems.
\end{abstract}

\maketitle

\section{Introduction}
Quantum Error Correction (QEC) plays a pivotal role in protecting qubits in the early fault-tolerant quantum era~\cite{terhal_quantum_2015,schindler_experimental_2011,chiaverini_realization_2004}. 
Quantum systems are inherently fragile, as they are highly susceptible to external disturbances such as noise, which can lead to decoherence and random bit flips, thereby compromising the integrity of quantum computations.
This makes QEC indispensable for reliable quantum computing, as it allows for the correction of such errors, ensuring the preservation of quantum information over time. Conventional QEC adapts the strategy of cyclic measurements of error syndromes and implementation of adaptive recovery operations, yet these methods come with several challenges~\cite{Nielsen_Chuang_2010, google2023suppressing, acharya2024quantum, gao2024establishing}. 

One significant issue with QEC is that the act of measuring qubits, in itself, can introduce errors due to the imperfection inherent in quantum measurements~\cite{campagne-ibarcq_quantum_2020,hu_quantum_2019,ofek_extending_2016,ma_error-transparent_2020,young_error_2013,atalaya_continuous_2021,chen_fast_2021}. This can lead to error propagation or even create new errors, thus paradoxically worsening the overall error rate~\cite{zhao_realization_2022,egan_fault-tolerant_2021}. Furthermore, the classical feedback loop requires extensive hardware resources, including additional qubits and complex control systems~\cite{fowler_surface_2012}. The hardware overhead poses a significant barrier to scale up quantum computers~\cite{terhal_quantum_2015}.

AQEC has emerged as a promising alternative, offering the advantage of bypassing the need for active measurements and recovery operations~\cite{gertler_protecting_2021,Barnes_automatic_2000,perez_improved_2020,steane_error_1996}. 
AQEC uses engineered interactions between qubits and ancillary systems, such as dissipative environments, which allows the entropy associated with decoherence to be transferred from the qubits to the ancillary system, where it can dissipate harmlessly into the environment~\cite{lihm_implementation-independent_2018,zanardi_dissipative_2016,hillmann_quantum_2023,cai_bosonic_2021,krastanov_room-temperature_2021,reiter_dissipative_2017,kerckhoff_designing_2010,kapit_hardware-efficient_2016}. 
This self-healing mechanism reduces the risk of introducing additional errors and minimizes hardware requirements.

Bosonic systems, with infinite-dimensional Hilbert space, provide an ideal platform for implementing AQEC codes. The well-defined noise channels in these systems make them particularly suited for robust error correction strategies. 
However, implementing these codes in practice is challenging. There are many complicated superpositions of the Fock states involved in these codes, which not only make the codes hard to generate, but also require high-order nonlinearity for gates to bridge between different Fock states.

To overcome the practical limitations of exact AQEC, researchers have turned to approximate AQEC~\cite{faist_continuous_2020,kong_near-optimal_2022,beny_perturbative_2011,beny_general_2010,leung_approximate_1997}, which relaxes the stringent Knill-Laflamme (KL) condition and allows for a broader, more implementable range of error correction codes. 
Recent advancements in artificial intelligence, particularly in deep reinforcement learning (DRL), have opened new avenues for optimizing QEC codes~\cite{fosel_reinforcement_2018,baum_experimental_2021,bolens_reinforcement_2021,xiao2022parameter, xiao2022intelligent, xiao2023quantum,webster2025engineering}.
By treating the search for optimal encoding strategies as a game, reinforcement learning (RL) algorithms can learn to identify code structures that maximize error correction performance while minimizing hardware requirements ~\cite{su2023discovery,zeng_approximate_2023,Mauron_2024,olle_simultaneous_2024,freire_optimizing_2025,su_discovery_2023,zeng_neural-network-based_2025}. 

Previous studies ~\cite{su2023discovery,zeng_approximate_2023} have successfully identified optimal encodings tailored to specific sets of system parameters, including cavity and qubit decoherence rates, along with coupling strengths. However, these encodings often lack adaptability when practical applications present varying or unforeseen conditions, underscoring the need for universally applicable encoding strategies capable of adjusting efficiently to diverse parameter sets. Additionally, second-order photon loss, which commonly occurs in bosonic systems at rates ranging from 1\% to 10\% of the single-photon loss rate ($\gamma_a$) ~\cite{tosca2024emergent}, has demonstrated significant detrimental effects on the performance of established encoding schemes such as T4C ~\cite{gertler_protecting_2021}, Binomial ~\cite{hu_quantum_2019}, and RL codes ~\cite{zeng_approximate_2023}. Consequently, exploring more sophisticated double-photon loss scenarios is imperative. 

Furthermore, these traditional methods typically focus on fidelity optimization at a single point in time or over short durations. While we build on the pioneering work of Ref. ~\cite{zeng_approximate_2023}, which first introduced reinforcement learning to tackle this complex, non-convex problem, their approach still struggles with generalization. In this work, we take a significant step forward by applying curriculum learning to enhance the reinforcement learning agent's ability to solve a broader range of problems, thereby addressing the challenges of long-term quantum evolution.

To address the challenges posed by increasingly large error spaces, conventional methods that employ simple projection operators for direct error correction become inefficient and impractical. In response, we adopt a cascading error-correction approach ~\cite{sivak_real-time_2023}, which incrementally projects error subspaces onto progressively closer subspaces to the original encoding space. This strategy avoids complex, experimentally challenging nonlinear operations. Furthermore, we integrate a dimension-reduction technique based on the master equation, enabling efficient and accurate simulations of quantum encoding performance. Prior research ~\cite{wang_automated_2022, li_autonomous_2023} primarily focused on addressing single-photon loss scenarios within specific system configurations. Hence, there remains a substantial opportunity and necessity for developing flexible, broadly applicable encoding strategies that can robustly handle diverse error mechanisms, significantly contributing to advancements in autonomous quantum error correction (AQEC).

In this paper, we first explore how second-order photon loss affects the fidelity of general encoding. Subsequently, by analyzing the KL conditions under photon-loss channels, we propose a simplified and universal RL model. We introduced curriculum learning (CL) to train the RL agent phase by phase efficiently. This model dynamically determines the current optimal encoding strategy based on system parameters and historical experiences, and, in certain cases, surpasses previous static encoding strategies. Finally, we conduct noise analyses of optimal encodings in several scenarios and provide methods for their practical implementation.

\section{Physical Model\label{sec:phy_mod}}
\subsection{Approximate AQEC\label{sec:physical_model-physics_system}}
\begin{figure}[!ht]
    \centering
    \includegraphics[width=8cm]{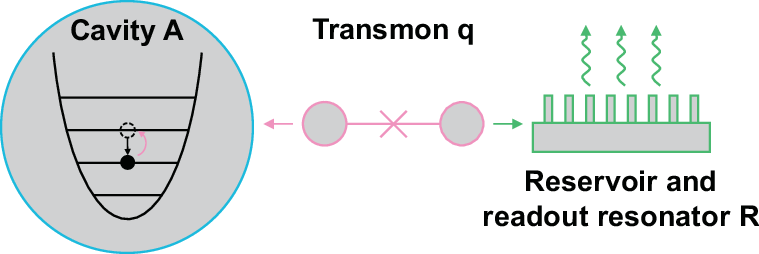}
    \caption{Schematic diagram of a typical AQEC system, including storage cavity $A$, transmon ancilla $q$ and readout $R$.}
    \label{fig:physics_layout}
\end{figure}

As shown in \Fref{fig:physics_layout}, the AQEC system generally consists of three main components: a system to be protected (denoted as $A$), an ancillary qubit (denoted as $q$) and a readout system (denoted as $R$). In this setup, errors and decoherence occurring in the primary system $A$ are transferred to the ancilla $q$, which then decays into the environment. The process of decoherence is thus offloaded from the primary system and dissipated by the ancilla. This passive correction mechanism is embedded directly into the system's dynamics, effectively "self-healing" the qubit states without the need for active control.

\begin{figure}[!ht]
    \centering
    \includegraphics[width=7cm]{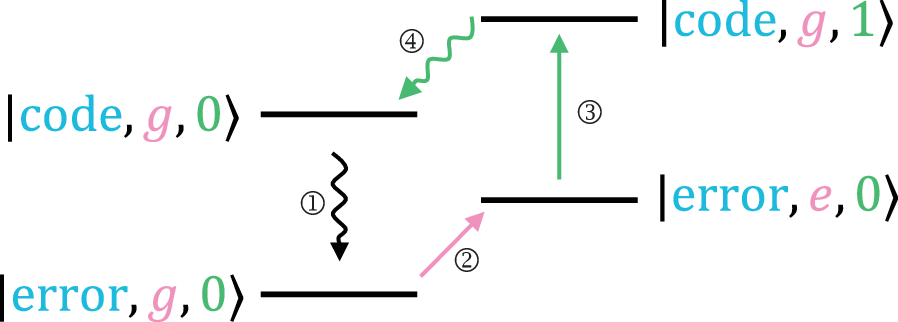}
    \caption{Hardware-module energy level diagram of approximate AQEC process. This diagram illustrates the physical "pump-and-dump" mechanism for a single recovery cycle. The labels 1-4 depict the error and recovery process of an AQEC code.}
    \label{fig:AQECstate}
\end{figure}

For the state $\ket{A,q,R}$, the AQEC process can be summarized as follows, as depicted in \Fref{fig:AQECstate}: 
\begin{enumerate}
    \item An error happens (arrow 1):\\
    $\ket{\mathrm{code},g,0}\rightarrow\ket{\mathrm{error},g,0}$
    \item The AQEC pump drives the transition (arrow 2 and 3):\\
    $\ket{\mathrm{error},g,0}\rightarrow\ket{\mathrm{error},e,0}\rightarrow\ket{\mathrm{code},g,1}$
    \item The readout R decays, returning the state to code space (arrow 4):\\  
    $\ket{\mathrm{code},g,1}\rightarrow\ket{\mathrm{code},g,0}$
\end{enumerate}
AQEC has shown considerable promise, with several codes developed to address the quantum error challenges. Codes such as the T4C code~\cite{gertler_protecting_2021}, binomial code~\cite{hu_quantum_2019} and $\sqrt{3}$ code~\cite{wang_automated_2022} fully satisfy the KL condition, theoretically enabling exact AQEC. However, the real-world implementation of these codes faces significant experimental challenges due to practical limitations.

In previous work~\cite{zeng_approximate_2023}, the KL condition was relaxed to allow for more flexible approaches that operate under a natural Lindblad operator $L_{\mathrm{nat,i}}\in\{I,a,a^2,\cdots\}$ by discarding the strict requirement that 
\begin{equation}
\bra{1_\mathrm{L}}L_{\mathrm{nat,i}}^\dagger L_{\mathrm{nat,i}}\ket{1_\mathrm{L}}=\bra{0_\mathrm{L}}L_{\mathrm{nat,i}}^\dagger L_{\mathrm{nat,i}}\ket{0_\mathrm{L}}    
\end{equation}
These RL-based codes have demonstrated superior performance compared to traditional AQEC codes, even surpassing the break-even threshold。
\subsection{Analytical solution of quantum master equation\label{sec:physical_model-analytical-solution}}

Currently, the search for practical encoding strategies is driven by classical simulations rather than experiments. However, the simulation process becomes time-consuming due to the computational bottleneck in solving the quantum master equation. 

To accelerate this process and enable the exploration of better encoding schemes, it is essential to develop more efficient yet still accurate methods. In this section, we introduce a new approach to analytically solve the quantum master equation for a typical AQEC system with appropriate approximations. Additionally, we describe how multi-photon losses can be included in this framework.

As discussed in Sec.\ref{sec:physical_model-physics_system}, engineered dissipation can be introduced through an ancillary qubit. The interaction is governed by the effective Hamiltonian: $H_{\mathrm{eff}}=g(L_{\mathrm{eng}}\sigma_{+}+L_{\mathrm{eng}}^{\dagger}\sigma_{-})$. Here, the engineered Lindblad operators $L_{\mathrm{eng}}$ projects the code space from error space. Consider a simple QEC jump operator with a Hamiltonian distance $d=1$
\begin{equation}
    L_o=\sum_{k\in\mathbb{K}}\lambda_k\ket{k+1}\bra{k}, \ \Lambda=\sum_{k\in\mathbb{K}}|\lambda_k|^2
\end{equation}
\begin{equation}
    L_{\mathrm{eng}}=L_{\mathrm{o}}/\mathrm{Tr}[L_{\mathrm{o}}^\dagger L_{\mathrm{o}}]=L_{\mathrm{o}}/\Lambda
\end{equation}
where $k\in\mathbb{K}$ denotes the set of chosen projectors. $\lambda_k$ is $0$ when the projector is not selected.
The dynamics of a typical AQEC process can be described by the master equation($\hbar=1$)~\cite{zeng_approximate_2023}.
\begin{equation}
    \frac{d\rho}{dt}=-i[H_{\mathrm{eff}},\rho]+\frac{\gamma_a}2\mathcal{D}[a]+\frac{\gamma_b}2\mathcal{D}[\sigma_-],
\end{equation}
The channel of dissipation is given by the Lindblad super-operator: $\mathcal{D}[x]=2x\rho x^\dagger-x^\dagger x\rho-\rho x^\dagger x$. $\gamma_a, \gamma_b$ refers to the rate of single-photon loss and auxiliary qubit decay, respectively. Taking the same assumption in ~\cite{zeng_approximate_2023} that $g, \gamma_a \ll \gamma_b$ and $\gamma_a\ll g$ are satisfied, the state can be transited uni-directionally from error space to code space. In this case, the density matrix of the hybrid system can be approximated as $\rho(t)=\rho_a(t)\otimes|0\rangle\langle0|$. Where $\rho_a(t)$ represents the state of encoding system, and $|0\rangle\langle0|$ is the ground state of auxiliary qubit. In this approximation, the effective dynamics described by the master equation act solely on the encoding subsystem~\cite{zanardi_dissipative_2016}, leading to the simplified equation:
\begin{equation}
    \frac{d\rho_a}{dt}=\frac{\gamma_a}{2}\mathcal{D}[a]+\frac{\gamma_a\lambda}{2}\mathcal{D}[L_{\mathrm{eng}}],\quad\lambda=\frac{8|g|^2}{\gamma_b\gamma_a}=8C.
\end{equation}
For each matrix element, the equation can be expressed as:
\begin{equation}
    \left(\frac{d\rho_a}{dt}\right)_{ij}=\frac{\gamma_a}{2}\mathcal{D}[a]_{ij}+\frac{\gamma_a\lambda}{2}\mathcal{D}[L_{\mathrm{eng}}]_{ij}, \ i,j=1,2\cdots
\end{equation}
Where the Lindblad operator $\mathcal{D}[a]_{ij}$ is given by:
\begin{equation}
    \mathcal{D}[a]_{ij}=2\sqrt{(i+1)(j+1)}\rho_{i+1,j+1}-(i+j)\rho_{i,j}
\end{equation}
For higher-order terms, $\mathcal{D}[a^n]$ follows a similar form:
\begin{equation}
\begin{split}
    \mathcal{D}[a^n]_{ij}=2\sqrt{A_{i+n}^nA_{j+n}^n}\rho_{i+n,j+n}
    -(A_{i}^n+A_{j}^n)\rho_{i,j}
\end{split}
\end{equation}
where $A_n^m=n!/(n-m)!$.
The Lindblad operator $\mathcal{D}[L_{\mathrm{eng}}]$ can also be simplified as:
\begin{equation}
    \mathcal{D}[L_{\mathrm{eng}}]_{i,j}=\frac{1}{\Lambda}\left[2\lambda_{i-1}\lambda_{j-1}\rho_{i-1,j-1}-(\lambda_i^2+\lambda_j^2)\rho_{ij}\right]
\end{equation}

Thus, we observe that $\rho_{i,j}$ is only related to $\rho_{i+n,j+n}$ in the master equation. This property allows us to decompose the equation into a set of decoupled equations for the diagonals of the density matrix.

\begin{figure}[h] 
  \centering
\begin{tikzpicture}[baseline=(m.center), rotate=53]
  \matrix (m) [matrix of math nodes,
    left delimiter=(,
    right delimiter=),
    nodes={inner sep=8pt, minimum size=1em}]
  {
    \rho_{11} & \rho_{12} & \rho_{13}\\
    \rho_{21} & \rho_{22} & \rho_{23}\\
    \rho_{31} & \rho_{32} & \rho_{33}\\
  };
  \coordinate (upperleft) at ($(m-1-1.north)+(-0.1cm,0.04cm)$);
  \coordinate (lowerright) at ($(m-3-3.south)+(0.1cm,0.0cm)$);
  \draw [red, thick, rounded corners=5pt] (upperleft) rectangle (lowerright);
  \coordinate (upperleft) at ($(m-1-2.north)+(-0.1cm,0.04cm)$);
  \coordinate (lowerright) at ($(m-2-3.south)+(0.1cm,0.0cm)$);
  \draw [orange, thick, rounded corners=5pt] (upperleft) rectangle (lowerright);
  \coordinate (upperleft) at ($(m-2-1.north)+(-0.1cm,0.04cm)$);
  \coordinate (lowerright) at ($(m-3-2.south)+(0.1cm,0.0cm)$);
  \draw [orange, thick, rounded corners=5pt] (upperleft) rectangle (lowerright);
  \coordinate (upperleft) at ($(m-1-3.north)+(-0.1cm,0.04cm)$);
  \coordinate (lowerright) at ($(m-1-3.south)+(0.1cm,0.0cm)$);
  \draw [yellow, thick, rounded corners=5pt] (upperleft) rectangle (lowerright);
  \coordinate (upperleft) at ($(m-3-1.north)+(-0.1cm,0.04cm)$);
  \coordinate (lowerright) at ($(m-3-1.south)+(0.1cm,0.0cm)$);
  \draw [yellow, thick, rounded corners=5pt] (upperleft) rectangle (lowerright);
  \node[above=0.6cm] at (m-1-1.center) {$\rho^{(0)}$};
  \node[above=0.6cm] at (m-1-2.center) {$\rho^{(1)}$};
  \node[above=0.6cm] at (m-1-3.center) {$\rho^{(2)}$};
  \node[left=0.8cm] at (m-2-1.center) {$\rho^{(-1)}$};
  \node[left=0.8cm] at (m-3-1.center) {$\rho^{(-2)}$};
\end{tikzpicture}
\caption{Definition of $\rho^{(m)}$ in the case of $N=3$}
\label{fig:diag_def}
\end{figure}
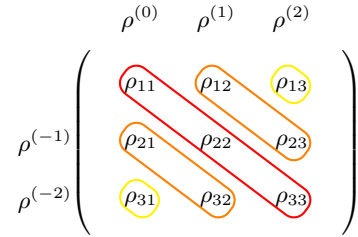

We define the $m$-th diagonal element ($m=-(N-1),\cdots,0,1,\cdots,N-1$ for an $N\times N$ tensor $\rho$) as: $\rho^{(m)}=[\cdots\rho_{i,i+m}\cdots]^T \ (i\mbox{ is interger for }1\le i,i+m\le N)$, as shown in \Fref{fig:diag_def}. 

The equation for each diagonal element can be written as:
\begin{equation}
\frac{d}{d(\gamma_\alpha t)}\rho^{(m)}=[\sum_{n=1}^N(\eta_n \mathcal{A}^{(m,n)})+\lambda\mathcal{L}^{(m)}]\rho^{(m)}
\end{equation}
where $\eta_n$ is the coefficient and the matrices $\mathcal{A}_{\alpha,\alpha}^{(m,n)}$ and $\mathcal{A}_{\alpha,\alpha+n}^{(m,n)}$ are defined as:
\begin{equation}
\begin{split}
\mathcal{A}_{\alpha,\alpha}^{(m,n)}=&-\frac{1}{2}(A^n_\alpha+A^n_{\alpha+m})\\
\mathcal{A}_{\alpha,\alpha+n}^{(m,n)}=&\sqrt{A^n_{\alpha+n}A^n_{\alpha+m+n}}
\end{split}  
\end{equation}
\begin{equation}
\begin{split}
\mathcal{L}^{(m)}_{\alpha,\alpha}=-\frac{\lambda_\alpha^2+\lambda_{\alpha+m}^2}{2\Lambda},\quad
\mathcal{L}^{(m)}_{\alpha-1,\alpha}=\frac{\lambda_{\alpha-1}\lambda_{\alpha+m-1}}{\Lambda}
\end{split}
\end{equation}
with all other elements being 0.

The solution to this system can be expanded as: $\rho^{(m)}(t)=\sum_{l}c_le^{w_lt}\cdot v_l$, where $w_l$ are the eigenvalues, $v_l$ the eigenvectors and $c_l$ are the corresponding coefficient. By assembling all the $\rho^{(m)}(t)$ back into $\rho_a(t)$, the master equation can be solved analytically.

\subsection{Reducing the Encoding Search Space}

To illustrate the advantage of this analytical solution, we consider a system with both single-photon loss and double-photon loss.
To balance both computational overhead and experimental feasibility, we truncate the Hilbert space at $N = 7$ Fock states. Nevertheless, allowing a fully unconstrained encoding remains challenging for RL, since the RL agent must maintain orthogonality within the code space. Our analytical consideration of the KL conditions indicates that $\langle 0_L \vert 1_L \rangle = 0$ implies an overlap of at least two Fock states if logical $\lvert 0\rangle$ and $\lvert 1\rangle$ share even one. Given that autonomous QEC encodings typically use at most three physical basis states, such a shared basis is impractical.  

Hence, we adopt a simplification wherein all Fock states are uniquely partitioned into the logical $\lvert 0\rangle$ or $\lvert 1\rangle$ subspace. This intuitive yet effective approach lowers the dimensionality of the RL action space from 15 (i.e., 8 coefficients for the logical 0 plus 7 for logical 1) down to 8, thereby reducing training complexity and improving overall performance. The encoding strategy can be expressed as:
\begin{equation}
    \ket{0_L}=\sum_{n\in \mathcal{H}_0}c_n\ket{n},
    \ket{1_L}=\sum_{m\in \mathcal{H}_1}c_m\ket{m}
\end{equation}

Building on the code space introduced above, $L_o$ is autonomously learned by the RL framework. However, in accordance with the technique above, we now restrict $L_o$ to lie within a linear regime.
\begin{equation}
    L_o=\sum_{n=1}^{7}d_n\ket{n}\bra{n-1}
\end{equation}

With this strategy, the dimension of encoding space increase linearly rather than quadratically with regard to the highest truncated Fock state $N$, which proves the scalability of this method.

In evaluating the performance of a QEC code, we define the mean fidelity $\bar{F}(\ket{0_{\mathrm{L}}},\ket{1_{\mathrm{L}}},t)$. Consider an arbitrary quantum state $\ket{\psi_{\theta\phi}}$ defined as:
\begin{equation}
    \ket{\psi_{\theta\phi}}=\cos{\frac{\theta}{2}}\ket{0_{\mathrm{L}}}+e^{i\phi}\sin{\frac{\theta}{2}}\ket{1_{\mathrm{L}}}
    \label{equ:quantum_state}
\end{equation}
where $\theta$ and $\phi$ represent the angles in Bloch sphere.
The density matrix corresponding to this state is given by $\rho_{t_0}(\theta,\phi)=\ket{\psi_{\theta\phi}}\bra{\psi_{\theta\phi}}$. After evolving for a time $t$, the state transforms into $\rho_{t}(\theta,\phi)$ and the fidelity between the initial state and the evolved state is defined as: $F(\theta,\phi,t)=\text{Tr}[\rho_{t_0}(\theta,\phi)\rho_{t}(\theta,\phi)]$. The mean fidelity $\bar{F}(t)$ represents the average of the fidelities over all possible states $\ket{\psi_{\theta\phi}}$ and is computed as:
\begin{equation}
    \bar{F}(t)=\frac{1}{4\pi}\int_\Omega F(\theta,\phi,t)d\Omega
    \label{equ:F_def}
\end{equation}
This integration can be simplified by expressing it as a sum over six basic density matrices. These six density matrices correspond to the Pauli matrices $\sigma_j$(where $j\in \{x,y,z\}$) ~\cite{zeng_approximate_2023}.
\begin{equation}
    \bar{F}(t)=\frac{1}{6}\sum_{j=\pm x,\pm y,\pm z}\text{Tr}(\rho_j(t_0)\rho_j(t))
\end{equation}
where $\rho_{\pm j}=(\sigma_0\pm\sigma_j)/2(j=x,y,z)$ and $\sigma_0$ the identity matrix.
\subsection{Curriculum reinforcement learning for AQEC}
\begin{figure*}[!ht]
    \centering
    \includegraphics[width=0.99\textwidth]{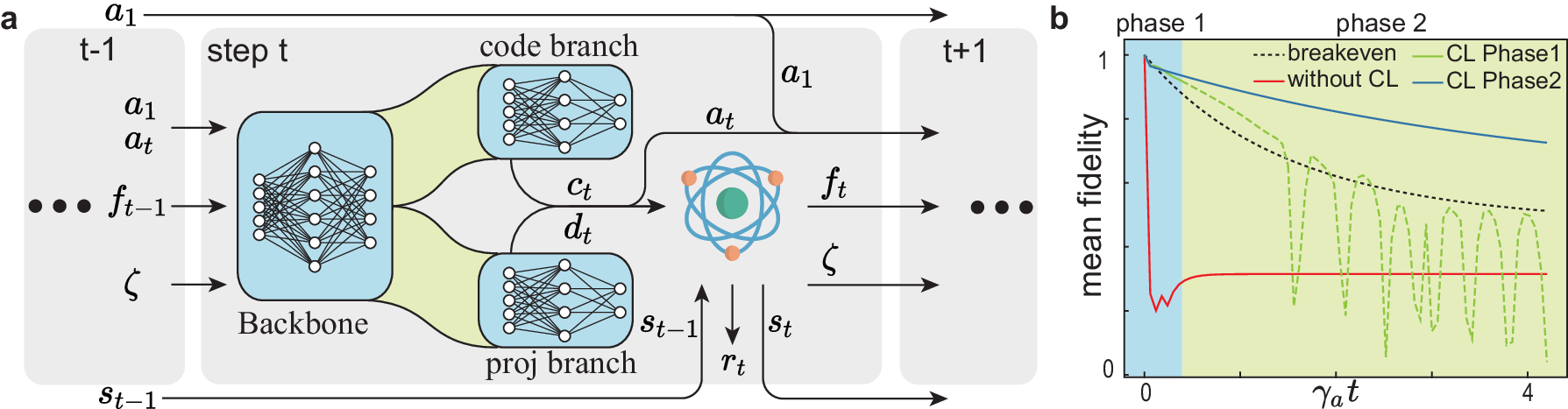}
    \caption{The diagram of training the PPO agent with the quantum environment. 
    \textbf{a} The training process of GRL code in each episode. The training of the agent lasts for 300k episodes. In each episode, the quantum environment first creates the six basic quantum states $s_1$ according to the codeword generated by its action $a_1$ in the first step, and takes the $s_{t-1}$ as the initial states in the following steps. These states evolve over time $dt=0.06/\gamma_a$, the density matrices $\rho_i,(i=1,\cdots,6)$ and the codeword are used to compute the fidelity $f_t=(\text{Tr}[\rho_{1,0}\rho_{1,t}],\cdots,\text{Tr}[\rho_{6,0}\rho_{6,t}])$. Based on $o_t=(f_t,a_t,a_1,\gamma_b/\gamma_a, \gamma_{a2}/\gamma_a,g/\gamma_a)$, the results are fed back into the agent, which determines a new codeword $a_{t+1}$. This process continues until the evolution time $t=4.2/\gamma_a$ at which point a new episode begins. Throughout the training, the agent gets a reward $r_t$ every action it takes based on the fidelity achieved. The agent's objective is to maximize the reward, which in turn helps it converge towards the optimal codeword. 
    \textbf{b} Comparison of RL performance with and without CL. Without CL, due to an excessively long simulation time $\gamma_at=4.2$, the initial steps' rewards diminish significantly because of the discount factor, making it challenging for the RL model to learn an initial density matrix that surpasses the breakeven fidelity. Consequently, RL fails to effectively enhance the overall reward (red curve). In contrast, by employing CL, we first restrict the simulation time of each episode to a shorter duration($\gamma_at\le 0.24$), encouraging the RL model to discover effective encodings exceeding the breakeven threshold (green curve). In the second phase, we gradually extend the maximum simulation duration; training continues as long as the average fidelity remains above the breakeven point. This strategy enables RL to explore increasingly stable encodings based on achievements in the first phase, sustaining fidelities above the breakeven level for longer periods (blue curve).
    }
    \label{fig:PPO_CL}
\end{figure*}

The RL frameworks are fundamentally built upon the formalism of Markov Decision Processes (MDPs), which are defined by a tuple $(S,A,P,R,\gamma)$, where $S$ is the set of states, $A$ is the set of actions, $P(s'|s,a)$ is the transition probability between states, $R(s,a)$ is the reward function, and $\gamma$ is the discount factor. The goal of the agent is to learn a policy $\pi(a|s)$, which maps states to a probability distribution over actions, so as to maximize the expected cumulative reward $\sum_{i=1}^N \gamma^i R_i(s,a)$ with $N$ denoting the maximum time steps in each episode.

In the context of AQEC, the exploration of effective encoding strategies and logical operations can be framed as a sequential decision-making problem, naturally modeled as an MDP. Here, the state can represent the six basic density matrices; the action corresponds to modifications of the logical qubit basis and Lindbladian operators; the reward is derived from metrics such as the improvement in logical fidelity over time, especially relative to the breakeven threshold. Given the total evolution time $t_i,i\in{1,2,\cdots N}$, the agent and MDP together decide a trajectory
\begin{equation}
    \langle S_{t_0},A_{t_0},R_{t_1},S_{t_1},A_{t_1},\cdots, R_{t_N},S_{t_N}\rangle
\end{equation}
The optimal code is to maximize the fidelity of all states in this trajectory. As shown in \Fref{fig:PPO_CL}a, the neural network architecture comprises a shared backbone and two output branches, which respectively provide the coefficients for the eight Fock states and for the seven projection operators. Specifically, each training iteration produces 8 coefficients for the Fock-state representation and 7 for the projection operators.

The model's input $o_t=[\tilde{a}_t,\tilde{f}_t,\tilde{\zeta}]$ includes three categories of features. First, it incorporates $\tilde{f}_t=[f_{t1},\cdots,f_{t6}]$, the fidelities of six reference states. Second, it uses normalized action values $\tilde{a}_t=[a_t,a_1]$ from the previous training step $a_t=[c_t,d_t]$ (15 parameters) and the initial training step $a_1=[c_1,d_1]$ (another 15 parameters). Third, it takes three system parameters—$\tilde{\zeta}=[\gamma_b/\gamma_a,\gamma_{a2}/\gamma_a,g/\gamma_a]$—which are each constrained within $[600,1800]$,$ [0,0.08]$, and $[300,600]$, respectively, aligning with ranges feasible in mainstream experimental platforms. During a single episode, these three system parameters remain fixed, but each episode begins with new random values to broaden the model’s applicability.

Because the agent’s actions specify coefficients for eight Fock states, we constrain each coefficient to the range $(-1, 1)$. Coefficients with positive values are assigned to the logical $\lvert 0\rangle$ component, and those with negative values are assigned to the logical $\lvert 1\rangle$ component, followed by normalization. This mapping simplifies the learning task by allowing the model to efficiently distinguish between logical states.

We extended the action space for RL to explore larger dimensions compared to previous works, which results in a significantly higher output dimensionality for our neural network. In RL scenarios of this scale, using conventional training methods could lead to excessively long training times or, in some cases, failure to converge. Consequently, in \Fref{fig:PPO_CL}b, we opt for a CL approach to accelerate and improve the training process.

The CL involves two phases. In the first phase, we constrained the maximum simulation time to $\gamma_a t=0.24$, corresponding to number of training steps per episode $K \le 4$, guiding the RL agent to converge its outputs from the entire action space towards a subspace that reliably exceeds the breakeven point. At each step, we define $\epsilon = \overline{F}(t) - \overline{F}_{\mathrm{be}}(t)$ as the difference between the average fidelity and the breakeven fidelity. The reward function for the first phase is given by 
\begin{equation}
    r_1 = f_1 \epsilon
\end{equation}
where $f_1$ denotes the reward normalization factor. This reward encourages the RL agent to identify encodings that surpass the breakeven threshold. 
This phase corresponds to a physical scenario where the encoded qubit is only required to outperform the breakeven point within a relatively short coherence time. Such a simplified objective leads to an easier training task for the RL agent, while still playing a crucial role in guiding the search toward a suitable initial encoding strategy.

In the second phase, our goal shifts to identifying an encoding that not only exceeds the breakeven threshold but also remains stable over time. To achieve this, we extend the training steps to a maximum of $K = 70 $ and introduce an action consistency reward, 
\begin{equation}
    \alpha=\frac{a_{t-1}\cdot a_t}{|a_{t-1}|\cdot|a_t|}
\end{equation}
ensuring the agent maintains some degree of flexibility in the encoding choices. Specifically, when $\alpha > 0.97$, we set $\alpha = 1$ to preserve the agent's flexibility. The updated reward function becomes
\begin{equation}
    r_2 = f_1 \epsilon +f_2\alpha
\end{equation}
The $f_2$ refers to the reward normalization factor of the action consistency, which make the second term of the reward no larger than the first one, keeping the agent's priority on maximizing the fidelity. In reality, we choose the hyper-parameters as $f_1=250, f_2=2$. 

Furthermore, if the fidelity drops below the breakeven threshold at any point, we assign a penalty of $r_3 = -20$ and immediately terminate the episode. This penalty encourages the agent to optimize its encoding, ensuring it converges to solutions that consistently meet the required performance criteria.

\begin{algorithm}[H]
\label{alg:CL-GRL} 
\caption{CL of GRL code including both phases}
\begin{algorithmic}
\State \textbf{Phase 1:}
\State Initialize Policy network $\pi_\theta$ and Value network $V_\phi$
\For{$k=1$ to $N_\text{1}$}
    \State start time $t=0$, time interval $dt$
    \State max of training steps in Phase 1 $K_1$ 
    \State random system parameters $\tilde{\zeta}$
    \State initial fidelity $\tilde{f}_0=[1,\cdots,1]$
    \State initial action $\tilde{a}_0=[0,\cdots,0]$
    \State Initialize state $o_0=[\tilde{a}_0,\tilde{f}_0,\tilde{\zeta}]$
    \While{$t<K\cdot dt$}
        \State Action $a_t \sim \pi_{\theta_{old}}(o_t)$ 
        \State Fidelity $F_{t+1} \gets \text{AnalyticalSolver.eval}(a_t)$ 
        \State $\epsilon_t \gets F_{t+1}$
        \State $\alpha_t \gets (a_{t-1},a_t)$
        \State Reward $r_t = f_1\epsilon$
        \State New state $o_{t} \gets (o_t, a_t,F_{t+1})$ 
        \State Store $(o_{t-1}, a_t, r_t, o_t)$
        \State $t=t+dt$
    \EndWhile
    \State For each time step, compute the advantage estimate
    \For{$e=1$ to number of training epochs}
        \State Sample a mini-batch of transitions
        \State Compute the total loss
        \State Update network parameters $\theta$ and $\phi$
    \EndFor
    \State Update policy parameters: $\theta_{old} \leftarrow \theta$
\EndFor
\State \textbf{Phase 2:}
\State Take $\pi_\theta$ and $V_\phi$ in Phase 1 as initial networks
\State new training iteration number $N_2$
\State new max of training steps in Phase 2 $K_2$ 
\State new reward function $r_t = f_1\epsilon+f_2\alpha$
\State Repeat training process
\end{algorithmic}
\end{algorithm}

The experience $(o_{t-1},a_t,r_t,o_t)$ is stored into the Replay Buffer. Random samples from the replay buffer are used to update the neural network parameters and improve the agent's ability to maximize the rewards. The detailed algorithm can be found in Appendix \ref{sec:PPO_train}(see Ref. ~\cite{astarrf_github_nodate} for codes).

\section{Result\label{sec:result}}
\begin{figure*}[!htb]
    \centering
    \includegraphics[width=0.95\textwidth]{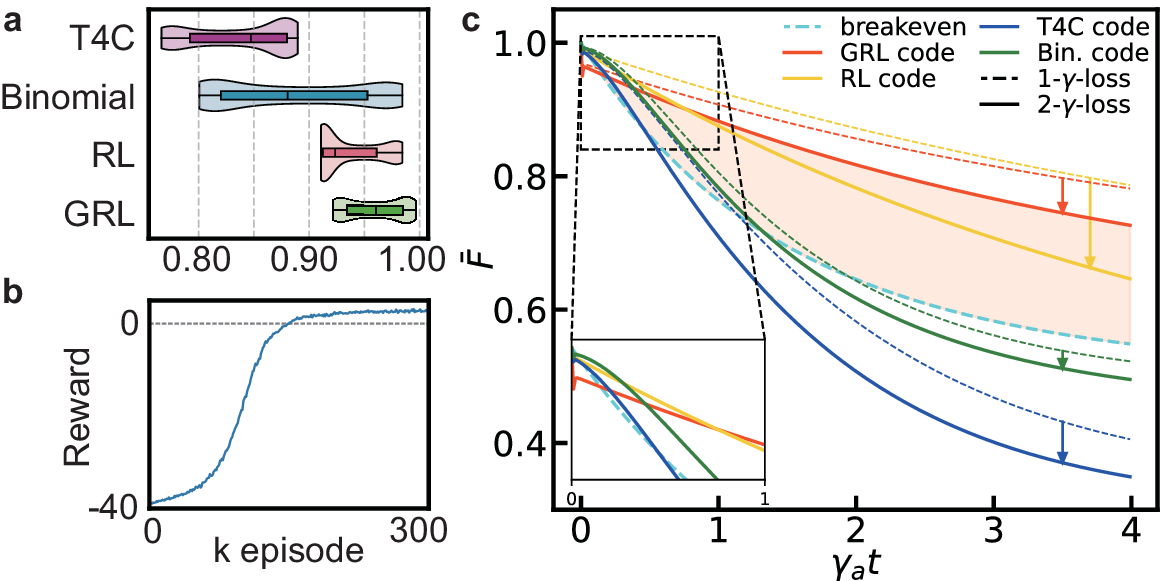}
    \caption{\textbf{a} The fidelity distribution $F(\theta,\phi,t)$ of other codes (T4C~\cite{gertler_protecting_2021}, Binomial~\cite{hu_quantum_2019} and RL~\cite{zeng_approximate_2023} code) and GRL code solved as the function of Bloch angle $\theta$ and $\phi$ analytically with step $\pi/10$ and $\pi/20$ respectively. The result is shown with $\gamma_at=0.6$ and $\lambda=10^4$.
    \textbf{b} Plot of average reward(solid blue line) as function of train episode. The evolution time for each step is $\gamma_a t=0.06$
    \textbf{c} Comparison of AQEC performance of: breakeven, T4C code, the lowest-order binomial code (Bin. code in figure), GRL code and RL code. Performance with same code are compared considering single-photon-loss(dash lines) and double-photon-loss(solid lines) to show the robustness versing higher-order-photon-loss.}
    \label{fig:compare_scan_reward}
\end{figure*}

The analytical master equation solver offers a significant advantage due to its lower algorithmic complexity. Solving the master equation numerically, without any simplification, would typically involve a single differential equation with $(2N)^2$ pivots, where $N$ represents the number of Fock states considered. This makes it computationally expensive. However, our algorithm takes the advantage of appropriate approximation and the sparsity of operators to reduce this complexity. The master equation is split into at most $2N-1$ differential equations, each having between $1$ and $N$ pivots. This decomposition leads to a significant reduction in computational cost, reducing the overall complexity by a factor of $O(N)$. This enables faster and more efficient solutions, making the approach scalable for larger systems. The tests presents in \Tref{tab:time_compare} demonstrate that this method achieves not only several orders of magnitude speed-up for larger matrix sizes, but also nearly twofold acceleration even at the relatively small scales required in this study. 
In our training, solving the master equations with QuTip can take approximately 73\% of the total training time. With the analytical solver, this process can be accelerated by 40\%.
\begin{table}[!htb]
\begin{ruledtabular}
\begin{tabular}{ccc}
N  & QuTip(s) & Analytical Solver(s)   \\ \hline
$8$  & $0.55$ & $0.30$ \\ 
$16$ & $2.82$ & $0.33$ \\ 
$32$ & $7.75$ & $0.37$ \\
\end{tabular}
\end{ruledtabular}
\caption{\label{tab:time_compare}Time for solving one typical AQEC code using both solvers under different scales of Fock states.}
\end{table}

The optimal code is denoted as the Generalized RL (GRL) code.
Considering even a small rate of double-photon loss $\gamma_{a2}=0.012\gamma_a$, this code significantly outperforms the previously established RL code, which under-performs in the presence of double-photon loss, as shown in prior work~\cite{zeng_approximate_2023}. The GRL code surpasses the breakeven, achieving a mean fidelity of $91\%$ when $\gamma_a t=0.6$. 
The optimal code corresponds to the states $\ket{0_{\mathrm{L}}}=\ket{4},\ket{1_{\mathrm{L}}}=\ket{7}$, 
with the engineering Lindblad operator $L_o\propto \ket{3}\bra{2}+\ket{4}\bra{3}+\ket{6}\bra{5}+\ket{7}\bra{6}$ (See Appendix. \ref{sec:rl_results} for details).
Remarkably, this result holds consistently for different evolution times $\gamma_a t$. 

The parameters used during testing reflect realistic experimental values. Specifically:
$g/\gamma_a=600,\gamma_b/\gamma_a=1800,\gamma_{a_2}=0.012\gamma_a$ 
where $g$ is the coupling constant, $\gamma_a,\gamma_{a_2}$ are the first-order and second-order decay rate of the cavity, and $\gamma_b$ represents the decay rate of ancilla.

The GRL code can be implemented in experiment quite conveniently. The QEC jump operator 
$L_{\text{eng}}\propto\ket{4}\bra{3}+\ket{3}\bra{2}+\ket{7}\bra{6}+\ket{6}\bra{5}$ has the shortest Hamiltonian distance $d=1$, preserving the key advantage of the RL code, which is that it can be implemented without the need for nonlinear interactions~\cite{gertler_protecting_2021}. For single-qubit gates, the Hamiltonian distance of GRL code is $d_g=3$, implying that third-order nonlinearity is required for its realization. For instance, the Pauli operators for the code can be:
$\sigma_x=\ket{4}\bra{7}+\ket{7}\bra{4},\sigma_y=i(\ket{4}\bra{7}-\ket{7}\bra{4}),\sigma_z=\ket{4}\bra{4}-\ket{7}\bra{7}$. In contrast, the T4C code and binomial code require fourth-order and sixth-order nonlinearity, respectively, with Hamiltonian distances of $d_g=6$ and $d_g=4$. This makes the implementation of these codes much more challenging.

\begin{figure*}[!ht]
    \centering
    \includegraphics[width=0.8\textwidth]{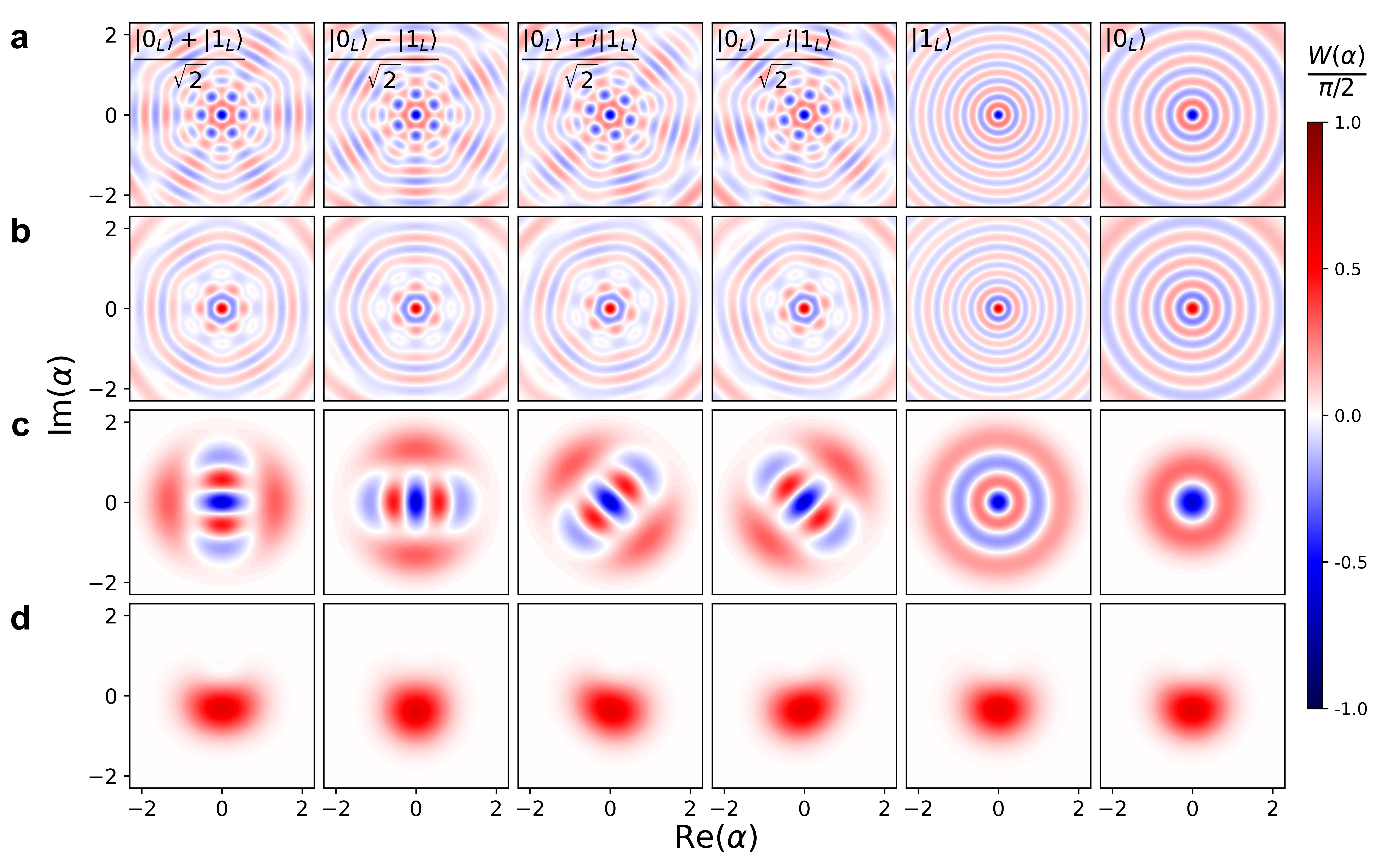}
    \caption{\textbf{a-d} Wigner function of six cardinal-point states of GRL code and breakeven before and after $\gamma_a t=4.2$ evolution.
    \textbf{a} GRL code before evolution.
    \textbf{b} GRL code after evolution with AQEC.
    \textbf{c} Breakeven before evolution.
    \textbf{d} Breakeven after free evolution.
    }
    \label{fig:wigner}
\end{figure*}

In \Fref{fig:compare_scan_reward}, we present a comparison of the GRL code's performance with breakeven and other well-known codes such as T4C code ($\ket{0_{\mathrm{L}}}=\sqrt{0.35}\ket{1}+\sqrt{0.65}\ket{5},\ket{1_{\mathrm{L}}}=\sqrt{0.9}\ket{3}+\sqrt{0.1}\ket{7}$), lowest-order binomial code ($\ket{0_{\mathrm{L}}}=\sqrt{0.5}\ket{0}+\sqrt{0.5}\ket{4},\ket{1_{\mathrm{L}}}=\ket{2}$) and RL code ($\ket{0_{\mathrm{L}}}=\ket{2},\ket{1_{\mathrm{L}}}=\ket{4}$). The performance comparison takes into account both single-photon loss (dashed lines) and double-photon loss (solid lines) to demonstrate the robustness of each code against higher-order photon losses. 
The GRL code consistently outperforms breakeven and all other codes in the presence of even a small probability of double-photon losses. While the fidelities of the other codes drop below breakeven, the GRL code maintains a mean fidelity above breakeven throughout the evolution. This superior performance is due to the unique error syndrome used by GRL code. Unlike other codes, which rely on photon number parity as the error syndrome ~\cite{ni_beating_2023}, the GRL code employs the$\mod{3}$ parity. This approach significantly enlarge the error space, with possible error states represented by: $\{\ket{0_{e1}}=\ket{3},\ket{1_{e1}}=\ket{6},\ket{0_{e2}}=\ket{2},\ket{1_{e2}}=\ket{5}\}$. Even with more syndrome, the GRL code can still use a correction operation with Hamiltonian distance $d_g=1$. Specifically, the correction operator can be: 

\begin{equation}
\begin{split}
L_{\text{eng}}\propto&\ket{4}\bra{3}+\ket{7}\bra{6}+\ket{3}\bra{2}+\ket{6}\bra{5}\\=&\ket{0_{\mathrm{L}}}\bra{0_{\mathrm{er1}}}+\ket{1_{\mathrm{L}}}\bra{1_{\mathrm{er1}}}\\
&+\ket{0_{\mathrm{er1}}}\bra{0_{\mathrm{er2}}}+\ket{1_{\mathrm{er1}}}\bra{1_{\mathrm{er2}}}
\end{split}
\end{equation}

These operators project the state from second-order error space to first-order error space, enabling the GRL code to handle double-photon loss with greater robustness. While it is not the optimal code for single-photon loss, the GRL code shows significant robustness to higher-order photon losses. The mean fidelity of GRL code drops by only $7.6\%$ in the presence of double-photon loss, compared to $21.7\%$ for RL code, $5.5\%$ for binomial code (which is lower than breakeven) and $16.1\%$ for T4C code. The suppression of fidelity loss shows the potential of cascaded error correction operators compares with conventional nonlinear ones.

The performance of the GRL code can also be better understood through the KL condition. The KL condition, which imposes two fundamental rules:
\begin{equation}
    \bra{0_\mathrm{L}}L_{\mathrm{nat,i}}^\dagger L_{\mathrm{nat,j}}\ket{1_\mathrm{L}}=0\label{equ:lgc_flp}
\end{equation}
and
\begin{equation}
    \bra{1_\mathrm{L}}L_{\mathrm{nat,i}}^\dagger L_{\mathrm{nat,j}}\ket{1_\mathrm{L}}=\bra{0_\mathrm{L}}L_{\mathrm{nat,i}}^\dagger L_{\mathrm{nat,j}}\ket{0_\mathrm{L}} \label{equ:dphs_free}
\end{equation}

The GRL code's performance is best explained by the Knill-Laflamme (KL) conditions for the error set $E = \{I, \sqrt{\gamma_a}\hat{a}, \sqrt{\gamma_{a_2}}\hat{a}^2 \}$. These conditions require (1) no logical-flip errors (\Eref{equ:lgc_flp}) and (2) a dephasing-free subspace (\Eref{equ:dphs_free}).

In the presence of double-photon loss, many bosonic codes fail the first rule because the $\hat{a}^2$ operator creates a direct logical-flip channel (i.e., $\langle 0_L | \hat{a}^2 | 1_L \rangle \neq 0$). This introduces a catastrophic error that scales with $\sqrt{\gamma_{a2}}$.

The key advantage of our GRL code ($|0_L\rangle = |4\rangle, |1_L\rangle = |7\rangle$)  is that its structure strictly satisfies this first rule. By ensuring $\langle 0_L | \hat{a}^2 | 1_L \rangle = 0$, it is immune to this logical-flip error. The GRL code does, however, violate the second rule (e.g., $\langle 0_L | \hat{a}^\dagger \hat{a} | 0_L \rangle \neq \langle 1_L | \hat{a}^\dagger \hat{a} | 1_L \rangle$). This violation introduces logical dephasing at a rate linear in the loss, but this is a far more manageable error.

Therefore, the GRL code's robustness stems from its design, which converts a catastrophic logical-flip error into a manageable dephasing error. This strategy explains its superior ability to maintain high fidelity in the presence of double-photon loss.

For codes that are structured in a similar form to $\ket{m},\ket{m+3}$, they are able to meet the KL condition more closely. However, as $m$ increases, these codes experience higher error rates due to the increased likelihood of photon loss events. In contrast, the GRL code serves as a balanced trade-off, offering optimal performance by minimizing the KL condition violation while also keeping the error rates in check. This makes the GRL code a preferable choice when considering the interplay between error tolerance and adherence to the KL condition.

It is important to note that there is an initial dip in fidelity, during which the GRL code performs worse than breakeven and the other codes. This short transition period has been discussed thoroughly in previous work~\cite{zeng_approximate_2023}. During this period, the jump operators are not effective because the code is entirely within code space. The decay rate is proportional to the mean photon number of the code. For the RL, binomial, T4C and breakeven codes, the mean photon numbers are $\bar{n}=3,2,3.5,0.5$,respectively, whereas for GRL code, the mean photon number is $\bar{n}=5.5$ . This fidelity dip represents reversible faults, which can be corrected using quantum trajectory-resolved and temporally coarse-grained fidelity $F^*_\tau(t)$~\cite{Sarma_2013}. Given that $\lambda=8|g|^2/\gamma_a\gamma_b\gg1$, the dip will become negligible~\cite{zeng_approximate_2023}.
\begin{figure*}[!ht]
\centering
    \includegraphics[width=0.9\textwidth]{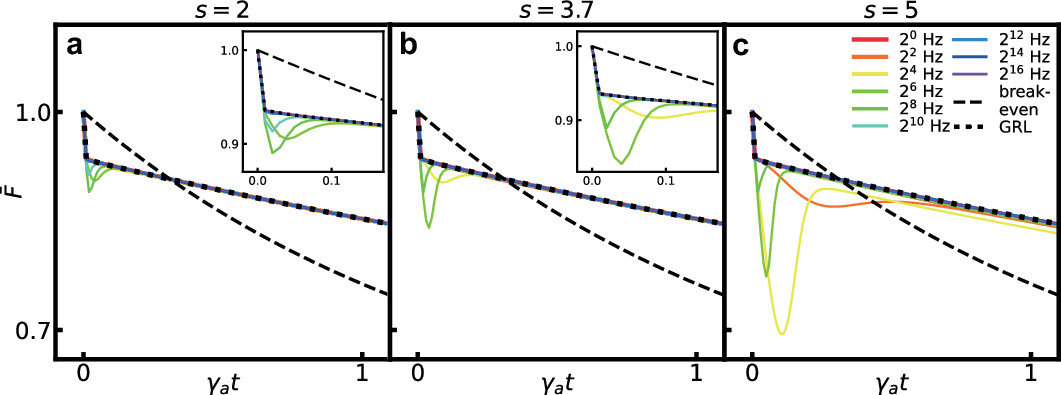}
    \caption{\textbf{a-c} The fidelity comparison between different $\omega_c$ with $s=2$ (a), $s=3.7$ (b) and $s=5$ (c). The phase damping noise is determined by 2 factors: $\omega_c$ and $s$. $\omega_c$ is the reservoir cutoff frequency and $s$ is the ohmicity parameter.}
    \label{fig:phase_damping}
\end{figure*}

The performance of GRL code is further illustrated using Wigner tomography. Six cardinal-point states are compared at two different time points: the initial states (\Fref{fig:wigner}a: GRL code; \Fref{fig:wigner}c: breakeven) and the states evolved until $\gamma_a t=4.2$ (\Fref{fig:wigner}b: GRL code; \Fref{fig:wigner}d: breakeven). As shown in these figures, the GRL code maintains a clearer and more distinct structure than the breakeven code, higher fidelity observed for the GRL code. Specifically, the mean fidelity for the GRL code is $F_{\text{GRL}}=70.5\%$, while for the breakeven, it is $F_{\text{breakeven}}=54.8\%$.

To further evaluate the consistency of the GRL code in protecting different quantum states, we simulated the evolution of all possible states in the code space, taking steps of $\Delta\theta=\pi/10,\Delta\phi=\pi/20$ at a time $t=0.6/\gamma_a$. The $\theta$ and $\phi$ here are the Bloch-sphere angles, as defined in \Eref{equ:quantum_state}. The resulting fidelities of the evolved states are shown in \Fref{fig:compare_scan_reward}a. Notably, the fidelity of all states in GRL code remains above $0.922$ , which means the GRL code outperforms the breakeven point $0.84$. When compared with other codes, the GRL code not only achieves the highest mean fidelity but also excels in both the maximum and minimum fidelity values. This indicates that the GRL code provides superior protection across the entire code space, maintaining a consistently high level of fidelity for all possible states.

To assess the robustness of the GRL code, we introduce two typical types of noise: phase damping and amplitude damping. A typical model for phase damping in a transmon system considers a well isolated cavity coupled to a transmon/ancilla, which interacts with a thermal bath. The master equation for this phase damping channel is described by the following Lindblad equation~\cite{dakir_quantifying_2024}:
\begin{equation}
    \frac{d\rho}{dt}=-i[H_{\mathrm{eff}},\rho]+\frac{\gamma_a}{2}\mathcal{D}[a]+\frac{\gamma_b}{2}\mathcal{D}[\sigma_-]+\frac{\gamma(t)}{2}\mathcal{D}[\sigma_z],
\end{equation}
The spectral density of the thermal bath is modeled as $J(\omega)=(\omega/\omega_c)^s e^{-\omega/\omega_c}$. where $\omega_c$ is the cutoff frequency and $s$ is the spectral index. The time-dependent damping rate $\gamma(t)$ is expressed as:
\begin{equation}
    \gamma(t)=\frac{\omega_c\Gamma[s]\sin{(\arctan{(\omega_c t)})}}{(1+(\omega_c t)^2)^{s/2}}
\end{equation}

For values of $s$ between $3.5$ and $4.5$, the noise transitions from Markovian to non-Markovian behavior, with the highest degree of non-Markovianity occurring around  $s\approx3.7$. We perform simulations for various values of $\omega_c$, focusing on $s=2,3.7$ and $5$, as shown in \Fref{fig:phase_damping}.The results indicate that, regardless of the Markovianity of the phase damping noise, the GRL code experiences an initial fidelity loss when $\gamma_a t<0.5$, with the most significant drop occurring at $\omega_c\approx2^8$Hz. However, the long-term fidelity loss remains negligible, staying below $<1.2\%$ even in the worst-case scenarios. This is because $\gamma(t)$ is nonzero only at the start. When the $\omega_c$ is too large, it quickly converges to zero before the initial dip ends, resulting in no noticeable effect on the fidelity. Conversely, if $\omega_c$ is too small, the amplitude of $\gamma(t)$ is insufficient to influence the fidelity. Phase damping becomes evident only when the $\omega_c$ is within an adequate range. With larger $s$, $\gamma(t)$ increases, leading to a deeper dip.
\begin{figure}[!ht]
    \includegraphics[width=0.48\textwidth]{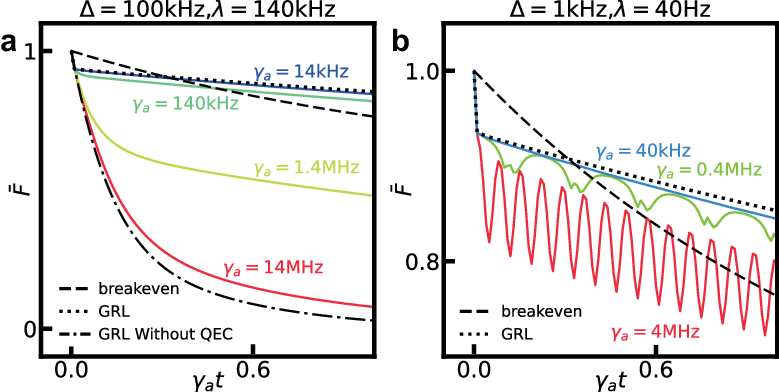}
    \caption{Fidelity evolution with amplitude damping. Results with different coupling coefficients are compared with breakeven, GRL code without noise and GRL code without both noise and error correction. The detuning frequency and width of Lorentzian distribution are $100$kHz and $140$kHz in (a) and $1$kHz and $40$Hz in (b) respectively.}
    \label{fig:amplitude_damping}
\end{figure}

Another significant noise source is amplitude damping, which occurs when a zero-temperature reservoir is coupled solely to the transmon. The dynamics of this noise are governed by the following master equation~\cite{dakir_quantifying_2024}:
\begin{equation}
    \frac{d\rho}{dt}=-i[H_{\mathrm{eff}}+H_{\mathrm{int}},\rho]+\frac{\gamma_a}{2}\mathcal{D}[a]+\frac{\gamma_b+\gamma(t)}{2}\mathcal{D}[\sigma_-]
\end{equation}
where the interaction Hamiltonian $H_{\mathrm{int}}$ is given by:
\begin{equation}
    H_{\mathrm{int}}=\frac{h(t)}{2}(\sigma_+\sigma_-)
\end{equation}
The time-dependent function $h(t)$ and damping rate $\gamma(t)$ are expressed as:
\begin{equation}
    h\left(t\right) =-2\mathrm{Im}\frac{\dot{\Re\left(t\right)}}{\Re\left(t\right)}
\end{equation}
\begin{equation}
    \gamma\left(t\right) =-2\mathrm{Re}\frac{\dot{\Re\left(t\right)}}{\Re\left(t\right)}=-\frac{2}{\left|\Re\left(t\right)\right|}\frac{d}{dt}|\Re\left(t\right)|
\end{equation}
The reservoir spectrum is described by a Lorentzian distribution:
\begin{equation}
    J\left(\omega\right)=\frac{\gamma_0\lambda^2}{2\pi\left[\left(\omega-\omega_c\right)^2+\lambda^2\right]}
\end{equation}
where $\gamma_0$ represents the coupling coefficient between reservoir and transmon, $\omega_c$ is the central frequency and $\lambda$ is the width of Lorentzian distribution. The function $\Re(t)$ evolves according to:
\begin{equation}
    \mathfrak{R}\left(t\right)=e^{\frac{-\left(\lambda-i\Delta\right)t}{2}}\left[\cosh\left(\frac{\Omega t}{2}\right)+\frac{\lambda-i\Delta}{\Omega}\sinh\left(\frac{\Omega t}{2}\right)\right]
\end{equation}
where $\Omega=\sqrt{\left(\lambda-i\Delta\right)^{2}-2\gamma_{0}\lambda}$ and $\Delta=\omega_0-\omega_c$ is the detuning between reservoir and transmon, with $\omega_0$ representing the transmon/ancilla frequency.

The fidelity of the system evolves differently depending on the detuning frequency at different scales. \Fref{fig:amplitude_damping} shows two typical cases. In the first case where $\gamma(t)\ll\gamma_b$, the interacting Hamiltonian $H_{\mathrm{int}}$ dominates the dynamics. When $\gamma_0$ is sufficiently large, $h(t)\gg g$, leading to the transmon degradation, and the fidelity approaches that of the non-QEC case. In contrast, when $\gamma(t)\gg \gamma_b$ and $h(t)\ll g$, the effective damping frequency of transmon is oscillating, resulting in fluctuation fidelity behavior.

The GRL code can be implemented without nonlinear interactions. For the setup in Sec.\ref{sec:phy_mod}, which may consists of a 3D coaxial cavity, a transmon qubit and a readout cavity $\mathcal{C}$ with fast decay rate $\kappa/2\pi=0.58$MHz, the Hamiltonian of the system is:
\begin{equation}
\begin{split}
        \frac{H}{\hbar}&=\omega_a a^\dagger a+\frac{\omega_b}{2}\sigma_z+\omega_c c^\dagger c+ \frac{\chi}{2}a^\dagger a\sigma_z\\
        &+ \chi_a(t)(a+a^\dagger)\sigma_x+\chi_c(t)(c+c^\dagger)\sigma_x
\end{split}
\end{equation}
$\omega_a$, $\omega_b$, $\omega_c$, are the resonant frequency of the cavity, transmon and readout respectively. To adiabatically eliminate the dynamics in transmon and readout, the rates follow $\omega_a,\omega_b,\omega_c\gg\chi$, and $\gamma_{c}\gg g_1\ge |g_0|\gg\gamma_{a},\gamma_{b}$. The decay rates are denoted as $\gamma_i(i=a,b,c)$. $\chi_a(t)$ and $\chi_c(t)$ here are time-dependent control fields. 
\begin{equation}
    \chi_a(t)=2g_0\sum_{n=3,4,6,7}{\frac{1}{\sqrt{n}}\cos{\left[(\omega_a+\omega_b+\frac{n-1}{2}\chi)t\right]}}
\end{equation}
\begin{equation}
    \chi_c(t)=2g_1\sum_{n=3,4,6,7}{\cos{\left[(n\chi)t\right]}}
\end{equation}
The $g_0$ and $g_1$ are the control field strength. To selectively transfer the energy from qubit to the readout mode $\mathcal{C}$, the condition $|g_0(t)|\le|g_1(t)|$ must be satisfied. Under the rotating wave approximation, the drive Hamiltonian becomes
\begin{equation}
    H_{\text{eff}}=g_0(L_o\sigma_++L_o^\dagger\sigma_-)+g_1\sum_{n=3,4,6,7}\ket{n}\bra{n}(c^\dagger\sigma_-+c\sigma_+)\label{equ:H_eff}
\end{equation}
\begin{figure}[!ht]
    \includegraphics[width=0.48\textwidth]{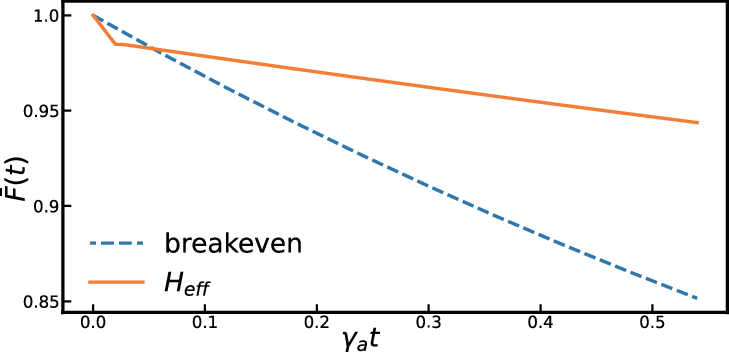}
    \caption{The fidelity of GRL code considering the coupling engineering. The parameters here are $\gamma_a=0.2\times2\pi\text{kHz},\gamma_{a2}=2\times2\pi\text{Hz},\gamma_b=2\times2\pi\text{kHz},\gamma_c=0.24\times2\pi\text{MHz},g_0=0.12\times2\pi\text{MHz},g_1=0.16\times2\pi\text{MHz},\omega_a=3.5\times2\pi\text{GHz},\omega_b=\omega_c=5\times2\pi\text{GHz}$}
    \label{fig:H_eff}
\end{figure}

From the simulation with the $H_\text{eff}$ in \Eref{equ:H_eff}, the \Fref{fig:H_eff} shows the GRL code surpasses the breakeven, with the gain $G=(1-\bar{F}_{be})/(1-\bar{F})\approx2.64$ at $t=3$ms.

The analytical solution can give a clearer picture of the protection of logical qubits. Under the same limit in ~\cite{zeng_approximate_2023} that $\gamma_a,g\ll\gamma_b$, $\gamma_a\ll g$, and $\gamma_a t\ll1,\lambda\gg 24$, the approximate solution of the first eight Fock states of the storage cavity is
\begin{equation}
    \rho_a(t)\approx \begin{pmatrix}
    0 & \cdots & 0 & 0 & 0 & 0\\
    \vdots & \ddots & \vdots & \vdots & \vdots & \vdots\\
    0 & \cdots & \rho_{44}(0) & 0 & 0 & \rho_{47}(0)e^{-u\gamma_at}\\
    0 & \cdots & 0 & 0 & 0 & 0\\
    0 & \cdots & 0 & 0 & 0 & 0\\
    0 & \cdots & \rho_{74}(0)e^{-u\gamma_at} & 0 & 0 & \rho_{77}(0)\\
    \end{pmatrix}
\end{equation}
The $\rho_{ij}(0)$ represents the initial states. The protecting factor
\begin{equation}
    u=\frac{11}{56}-\frac{\sqrt{2}}{14}+\frac{27}{112}\frac{\gamma_{a2}}{\gamma_a}\approx(7.44+241.07\eta)\times10^{-3}, \eta=\frac{\gamma_{a2}}{\gamma_a}
\end{equation}
is smaller than $26.7\times10^{-3}$ even with the largest $\gamma_{a2}=0.08\gamma_a$. The mean fidelity is therefore able to be derived from \Eref{equ:F_def}

\begin{equation}
\begin{split}
    \bar{F}(t)=&\frac{1}{4\pi}\oiint \rho_{44}^2(0)+\rho_{77}^2(0)+2|\rho_{47}(0)|^2e^{-u\gamma_at} \ \mathrm{d}S\\
    =&\frac{2}{3}+\frac{1}{3}e^{-u\gamma_at}, u<26.7\times 10^{-3}
\end{split}
\end{equation}

\section{Conclusion\label{sec:conclusion}}
In summary, we discover a new Bosonic AQEC code that has the shortest Hamiltonian distance $d=1$ using a curriculum learning (CL)-enhanced deep reinforcement learning (DRL) framework, which considers higher-order photon losses and is adaptive to even large Fock space. This framework addresses higher-order photon losses through a two-phase training strategy: a rapid exploration phase identifies subspaces surpassing the breakeven threshold, followed by a refinement phase that optimizes encoding policies for long-term stability. The resulting code demonstrates strong robustness, outperforming conventional RL-based and existing AQEC codes even under non-negligible double-photon-loss conditions.

To overcome computational bottlenecks, we develop a semi-analytical solution that accelerates DRL training while maintaining alignment with full quantum master equation simulations. Remarkably, the code exhibits resilience to both phase and amplitude damping noise, maintaining fidelity above the breakeven threshold across realistic noise regimes. Specifically, it recovers near-noise-free performance after transient phase damping and remains stable under amplitude damping with moderate coupling strengths.

Our work establishes CL-enhanced DRL as a transformative paradigm for designing resource-efficient QEC codes. Future directions include extending logical qubit lifetimes through combinatorial energy-level engineering and scaling the framework to multimode bosonic systems. These advances pave the way for practical fault-tolerant quantum memories and modular quantum processors.

\section*{Acknowledgment}
This work was supported by the National Natural Science Foundation of China (No. 62401359, 62471289), Quantum Science and Technology-National Science and Technology Major Project (Grant No. 2021ZD0300703), Shanghai Municipal Science and Technology Major Project (2019SHZDZX01) and SJTU-Lenovo Collaboration Project (202407SJTU01-LR019).

\bibliography{AQEC1}
\appendix

\section{The effect of double-photon loss on the code fidelity }
\begin{figure}[!ht]
    \includegraphics[width=0.48\textwidth]{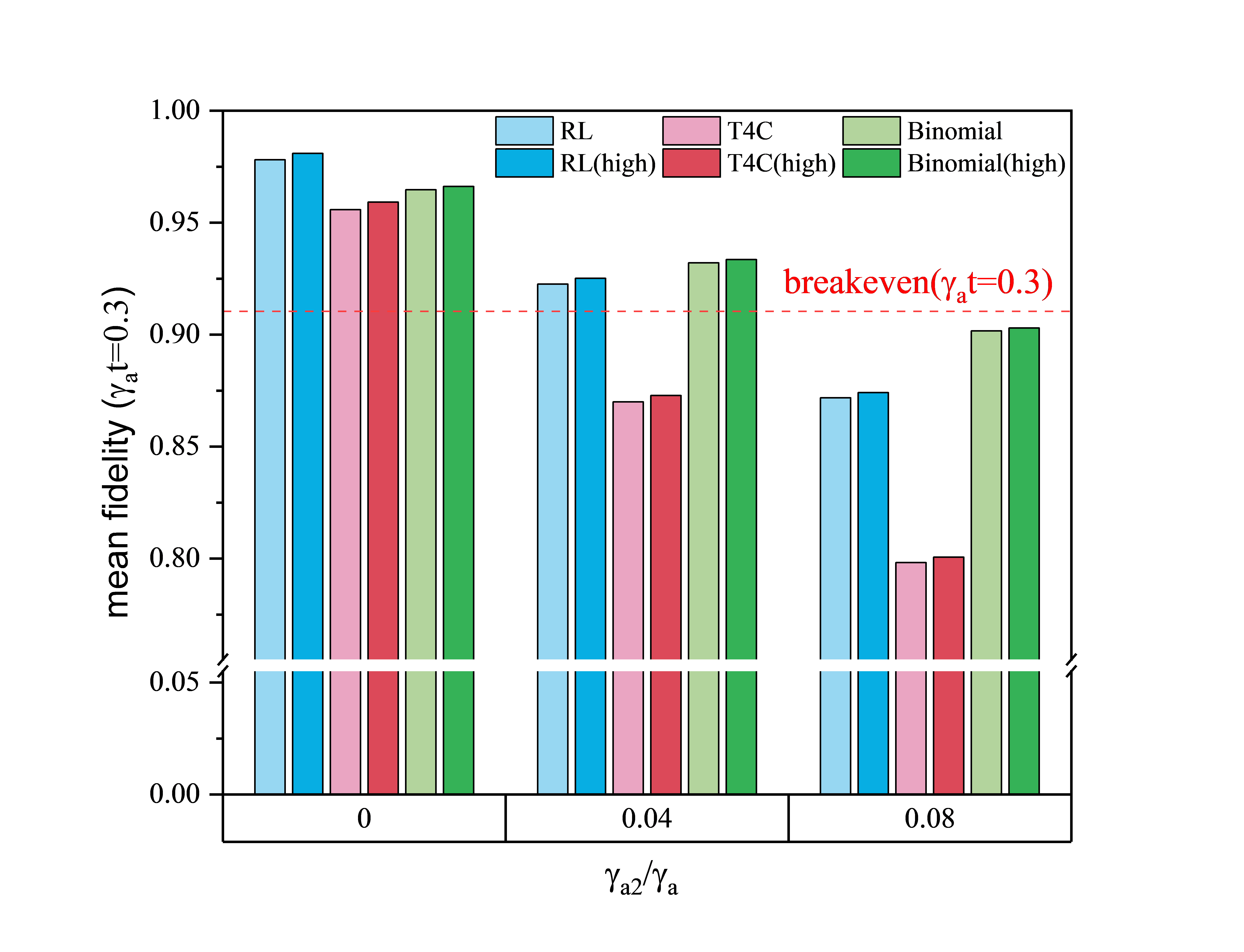}
    \caption{
    The mean fidelity of RL, T4C and Binomial codes under different second-order conditions ($\gamma_{a2}/\gamma_a=0,0.04,0.08$). The performance of each code is evaluated under the low coupling rate ($g/\gamma_a=600$) and high coupling rate ($g/\gamma_a=900$). The high coupling rate shows little levitation in the mean fidelity.
    }
    \label{fig:gamma_a2}
\end{figure}
To validate the nontrivial impact of second-order photon loss on fidelity, we evaluated how fidelity varies with different values of $\gamma_{a2}$ for several well-known codes. \Fref{fig:gamma_a2} shows the fidelity of each code at $\gamma_a t = 0.3$. For the RL, Binomial, and T4C codes, the mean fidelity already decreases by approximately 10\% at $\gamma_{a2}/\gamma_a = 8\%$, which is lower than the breakeven. This significant impact arises because the RL and Binomial codes exclusively perform parity checks for error correction. Under double-photon loss, the relationship $\bra{0_L}a^2 |1_L\rangle \neq 0$ emerges, causing a severe violation of the KL conditions when considering the error set $E = \{I, a, a^2\}$.

\section{PPO2 for searching the optimal code space\label{sec:PPO_train}}
Searching for the optimal states $\ket{0_L}$ and $\ket{1_L}$ is a complex optimization problem that we address using PPO. Each episode of training is separated into several steps $k=1,2\cdots K$. At step $k$, the network estimates the possible rewards $r_k$ for different actions $a_k\in A$, based on current status of the system $s_k\in S$. This process is described by the policy $\pi(A|S)$, which the network aims to optimize. In our case, the state $s_k$ includes the fidelities of six basic density matrices $\text{Tr}[\rho_j(t_0)\rho_j(t)]$, $j=\pm x,\pm y,\pm z$ which are calculated by the analytical solver. 
The action $a_k$ includes the coefficients of all the Fock states $c_n$ and the projection operators $d_n$. For the Fock states, the coefficients are restricted in the range $(-1,1)$, with the codeword
\begin{equation}
    \ket{0_L}=\sum_n \max{(c_n,0)}\ket{n}, \ \ket{1_L}=\sum_n -\min{(c_n,0)}\ket{n}
\end{equation}
In this way, both orthogonality of the codeword and the simplicity of the action space can be guaranteed.
The reward $r_k$ is designed to search for the code which is both robust and stable over time. We define 
$\epsilon_k=\bar{F}_k(\ket{0_L},\ket{1_L})-\bar{F}_{be}(\ket{0},\ket{1})$
as the fidelity gain of the current code compared to the breakeven fidelity.
Another part of the reward $\alpha=\frac{a_{t-1}\cdot a_t}{|a_{t-1}|\cdot|a_t|}$ is the action consistency. $\alpha$ is set to 1 when $\alpha>0.97$ in order to keep enough freedom of time-dependent codeword. The total reward is $r=50\epsilon$ in phase 1 and $r=250\epsilon+2\alpha$ in phase 2.

The code space is limited to the first 8 Fock state. Other parameters in the simulation are chosen as $g/\gamma_a=600,\gamma_b/\gamma_a=1800$ and $\gamma_a t=0.06$ for each of the steps per episode. 

The PPO algorithm involves 2 networks: the actor network and the critic network. The actor decides the actions according to its policy $\pi_\theta(a_t|o_{t-1})$, which consists of a backbone and two branches for the Fock states part and the projection operator part. The critic predicts the advantage $A_t$ of the next state, where
\begin{equation}
    A_t=-V(o_{t-1})+r_t+\gamma r_{t+1}+\gamma^2 r_{t+2}+\cdots
\end{equation}
When $A_t>0$, the agent estimates the current action is above the average, and encourages higher probability of this action.
$\gamma$ refers to the discount rate of the reward. The discount is set to 0.99 in most cases, ensuring the agent emphasis on the reward of current step, while still putting an eye on the future. The common loss function is:

\begin{equation}
    L(\theta)=\mathbb{E}_t\left[\log{\pi_\theta(a_t|o_{t-1})A_t}\right]
\end{equation}

However, this function often overestimate the policy update gradient empirically, leading to performance degradation. Therefore, we define a ratio $r_t(\theta)=\frac{\pi_\theta(a_t|o_{t-1})}{\pi_{\theta_{old}}(a_t|o_{t-1})}$ that evaluates the changes to the policy. The new loss function is~\cite{schulman2017proximal}:
\begin{equation}
    L^{CLIP}(\theta)=\mathbb{E}_t\left[\min{(r_t(\theta)A_t,\text{clip}(r_t(\theta),1-\epsilon,1+\epsilon)A_t)}\right]
\end{equation}
which restricts the ratio between $1-\epsilon$ and $1+\epsilon$. The $\epsilon$ here is a hyperparameter. 

\section{Comparison between QuTip and analytical solver in AQEC regime}
The training process of the GRL involves two main components: generating samples $(o_{t-1}, a_t, r_t, o_t)$ and updating the neural network based on these samples, executed separately on CPU and GPU, respectively. The former step requires solving the evolution of the density matrix. For the scenarios presented in this study, we truncate simulations at a maximum Fock state of $N=8$, as experimentally preparing higher-order Fock states remains highly challenging. During training, the simulation time accounts for roughly 73\% of the total training duration; hence, optimizing the solver becomes essential.

The Qutip library is commonly utilized for solving open quantum systems. Nevertheless, for specific QEC scenarios and in terms of computational efficiency, employing specialized and optimized solvers can yield superior performance. The model considered here involves coupling between a single mode and a dissipative mode, satisfying condition ($g, \gamma_a \ll \gamma_b$ and $\gamma_a\ll g$), and the engineered super-operator $L$ is represented as a linear combination of projection operators. As discussed in Section \ref{sec:physical_model-analytical-solution} of the main text, such a model can be simplified through several computational techniques, breaking down the full problem into smaller sub-equations and thus significantly reducing the computational complexity.

Here, we compare the computational time required to solve for fidelity variations of the same GRL code as a function of increasing maximum truncated Fock states used in the simulations. The QuTip takes longer time than the analytical solver in all circumstances. When the amount of involved Fock states increases, the Qutip shows an polynomial time consumption, while the analytical solver, which extracts and solves only the valid parts, only undergoes an linear increase in time.

\section{The approximation of numerical GRL results\label{sec:rl_results}}
\begin{figure}[!htb]
    \includegraphics[width=0.48\textwidth]{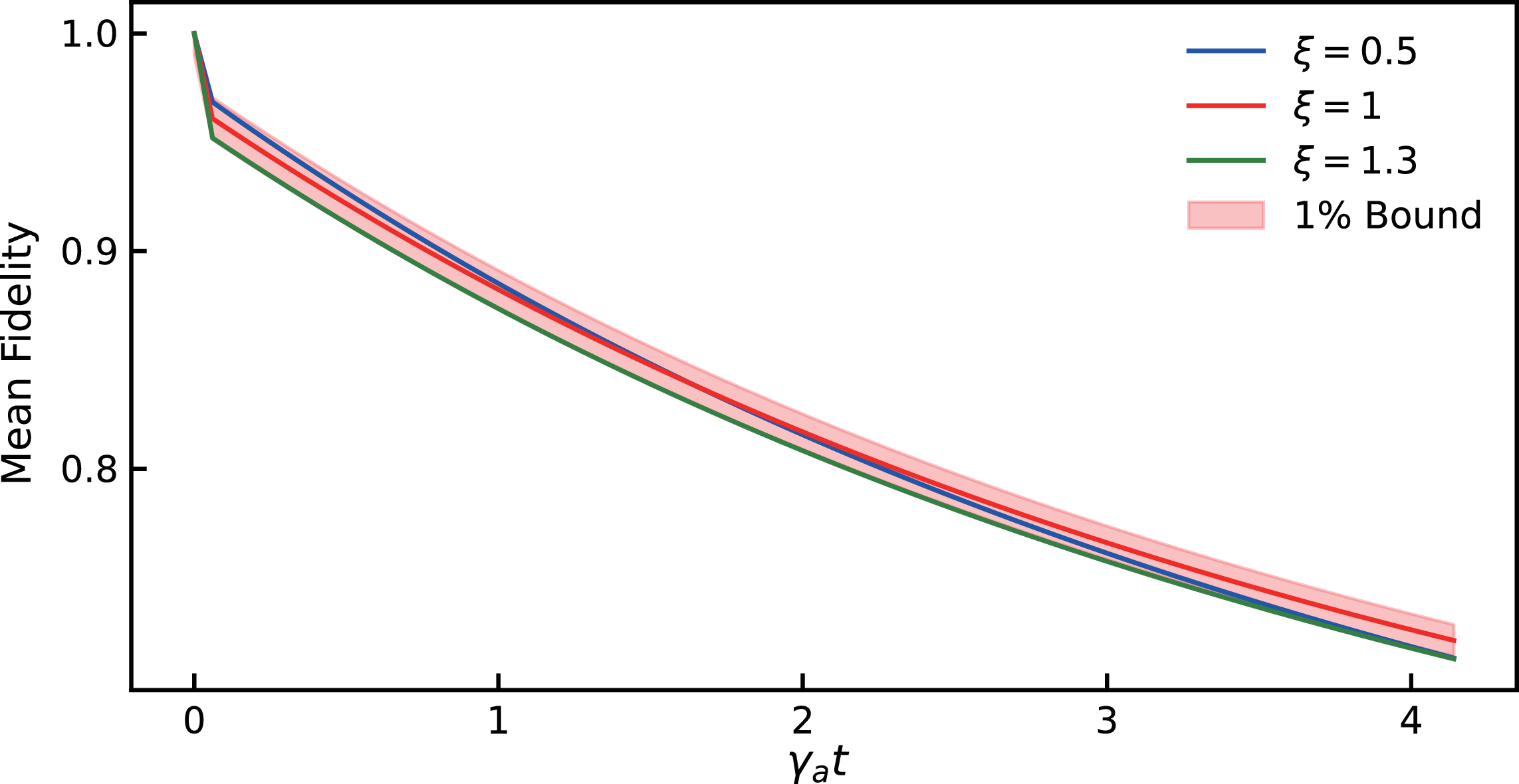}
    \caption{Mean fidelity as the function of $\gamma_at$ with $\xi=0.5,1.0,1.3$. The red shaded region represents the $\pm1\%$ error band surrounding the $\xi=1.0$ curve.}
    \label{fig:proj_analysis}
\end{figure}
The numerical results of the GRL code in \Tref{tab:GRL_ceof} is time-dependent with different coefficient for some projection operators. 

\begin{table}[!htp]
\begin{ruledtabular}
\begin{tabular}{ccccc}
step  & $\ket{3}\bra{2}$ & $\ket{4}\bra{3}$ & $\ket{6}\bra{5}$ & $\ket{7}\bra{6}$ \\ \hline
$1$&$0.50005746$&$0.50003510$&$0.49975090$&$0.50015640$\\
$2$&$0.49999446$&$0.50023013$&$0.50005066$&$0.49972460$\\
$3$&$0.50000340$&$0.50014806$&$0.50006340$&$0.49978513$\\
$4$&$0.50000370$&$0.50014406$&$0.50006360$&$0.49978864$\\
$5$&$0.50000410$&$0.50014466$&$0.50006490$&$0.49978620$\\
$6$&$0.50000450$&$0.50014470$&$0.50006590$&$0.49978480$\\
$7$&$0.50000490$&$0.50014480$&$0.50006700$&$0.49978337$\\
$8$&$0.50000530$&$0.50014480$&$0.50006800$&$0.49978197$\\
$9$&$0.50000566$&$0.50014480$&$0.50006896$&$0.49978057$\\
$10$&$0.50000600$&$0.50014484$&$0.50006990$&$0.49977922$\\
$11$&$0.50000640$&$0.50014484$&$0.50007080$&$0.49977788$\\
$12$&$0.50000680$&$0.50014484$&$0.50007170$&$0.49977657$\\
$13$&$0.50000715$&$0.50014490$&$0.50007266$&$0.49977530$\\
$14$&$0.50000750$&$0.50014490$&$0.50007350$&$0.49977400$\\
$15$&$0.50000787$&$0.50014496$&$0.50007430$&$0.49977273$\\
$16$&$0.50000820$&$0.50014500$&$0.50007520$&$0.49977150$\\
$17$&$0.50000864$&$0.50014510$&$0.50007600$&$0.49977030$\\
$18$&$0.50000900$&$0.50014510$&$0.50007683$&$0.49976910$\\
$19$&$0.50000930$&$0.50014514$&$0.50007760$&$0.49976790$\\
$20$&$0.50000966$&$0.50014520$&$0.50007840$&$0.49976674$\\
\end{tabular}
\end{ruledtabular}
\caption{\label{tab:GRL_ceof}Normalized coefficient of the four main projection operators of GRL code in the first 20 steps.}
\end{table}
The coefficients are evolving through the time, which is challenging for real implementation. However, tests in \Fref{fig:proj_analysis} shows that considering the form of the operator $L_o\propto \ket{4}\bra{3}+\ket{7}\bra{6}+\xi(\ket{3}\bra{2}+\ket{6}\bra{5})$. The variance of mean fidelity is smaller than 1\% within the range of $\xi\in[0.5,1.3]$. The result not only guarantee the simplification of the Lindblad operator in GRL code, but also exhibits the robustness of the cascading error-correction method.

\section{Detailed explanation of the alternative RL reward design}
\begin{figure}[!htp]
	\centering
	\includegraphics[width=0.48\textwidth]{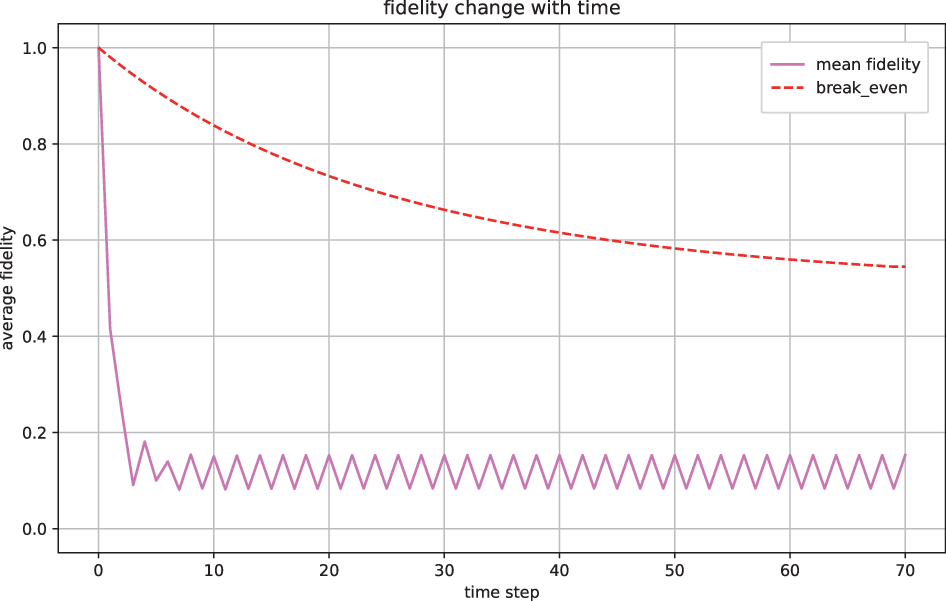}
	\caption{The mean fidelity as the function of time step. Each step is $\gamma_at=0.06$ long.}
	\label{fig:fid_curve_app}
\end{figure} 
An alternative reward structure based on the incremental fidelity improvement can be $\Delta F_k = F_k - F_{k-1}$, which rewards the RL agent for fidelity gains while penalizing declines. Although intuitively appealing, this approach was found ineffective in practice. Specifically, when implemented, the RL agent quickly learned to artificially generate incremental rewards by causing oscillations in the fidelity values without achieving sustained meaningful improvement, as demonstrated in ~\Fref{fig:fid_curve_app}. Such oscillatory behavior undermined effective training and optimization.
\end{document}